\documentclass[letter,nolinenumbers]{aa}  
\usepackage{xspace}
\usepackage{graphicx}
\usepackage{txfonts}
\usepackage{lipsum}
\usepackage{subcaption}         % necessary for continued figures, example in section 3
\usepackage{lscape}             % to rotate a single page table, example in appendix.
\usepackage{placeins}           % useful with \FloatBarrier, to keep 
\usepackage{hyperref}
\hypersetup{colorlinks,%
 citecolor=blue,%
 linkcolor=blue}
 
\usepackage{linenoaa}

 \newcommand{\kms}{${\rm km\,s^{-1}}$\xspace}
\newcommand{\isde}{dE01+09\xspace}
  
\begin{document}

   \title{ An isolated early-type dwarf galaxy that ran away from the group environment}

   \subtitle{}

   \author{Sanjaya Paudel\inst{1,2}
        \and Cristiano G. Sabiu\inst{3}
          \and Suk-Jin Yoon \inst{1,2}
          \and Patrick R. Durrell \inst{4}
         \and Nau Raj Pokhrel \inst{5}      
        %\fnmsep\thanks{}
        }

   \institute{Department of Astronomy, Yonsei University, Seoul 03722, Republic of Korea\
             \email{sjyoon0691@yonsei.ac.kr}
             \thanks{Shows the usage of elements in the author field}
            \and Center for Galaxy Evolution Research, Yonsei University, Seoul 03722, Republic of Korea 
            \and Natural Science Research Institute (NSRI), University of Seoul, Seoul 02504, Republic of Korea
            \and Department of Physics, Astronomy, Geology and Environmental Sciences, Youngstown State University, Youngstown, OH, USA
            \and Department of Physics and Astronomy, The University of Tennessee, Knoxville, TN 37996, USA
            }

   \date{Received September 30, 20XX}

\abstract{
Understanding the quenching mechanisms in dwarf galaxies is crucial for constraining models of galaxy formation and evolution. In this vein, isolated dwarf galaxies offer valuable insight by helping disentangle the relative roles of internal and environmental processes in shutting down star formation. Here we report the discovery of a quiescent early-type dwarf galaxy (dE), SDSS J011754.86+095819.0 (hereafter \isde), located in a nearly isolated environment at a projected distance of approximately one megaparsec from its most likely host group, the NGC\,524 group. \isde has $M_r$\,=\,$-15.72$ and $g-r$\,=\,0.67 and its light profile is well described by a S\'ersic function with an index $n$\,=\,1.1, consistent with typical dEs.
Using optical spectroscopy from the DESI survey, we derive its simple stellar population properties, finding an intermediate luminosity-weighted age of 8.3\,$\pm$\,1.4\,Gyr and a subsolar metallicity of $-1.19$\,$\pm$\,0.21\,dex---characteristics comparable to those of classical quiescent dEs. We propose that NGC\,524 may represent an extreme example of group dynamics, in which a member galaxy, \isde, is ejected from its host group and subsequently evolves as an isolated system in the field.
}

\keywords{Galaxy:general -- galaxies:dwarf -- galaxies:environment }

\maketitle

%%%%%%%%%%%%%%%%%%%%%%%%%%%%%%%%%%%%%%%%%%%%%%%%%%%%%%%%%%%%%%
\section{Introduction}

Understanding how galaxies quench is essential for explaining the various internal and environmental processes that shape galaxy evolution. Quenching typically involves the depletion, removal, or disruption of a galaxy’s reservoir of star-forming gas. Internal factors such as active galactic nuclei, supernovae, and stellar feedback play a key role in halting star formation by heating or expelling cold gas \citep{Croton06, Hopkins12}. On the other hand, environmental processes such as ram-pressure stripping (RPS), tidal harassment, and starvation can also remove or deplete the gas reservoir \citep{Gunn72, Larson80, Moore96, Simpson18, Cortese21}.

The strong correlation between morphology and density observed in low-mass galaxies is widely believed to be shaped by environmental effects during their evolution. In particular, early-type dwarf galaxies (dEs)---a class of low-mass, non-star-forming galaxies---are most commonly found in galaxy clusters or groups, suggesting that their origins are closely tied to environmental influences \citep{Binggeli88, Boselli06, Lisker07}. Despite this, the relative contributions of different environmental mechanisms---such as RPS, harassment, or starvation---in shaping the stellar populations and structural features of dEs remain unclear \citep{Boselli06, Mayer01, Smith10}. Among these, RPS is considered especially effective in dense cluster environments, with numerous recent studies reporting active cases of RPS, particularly within the Virgo Cluster \citep{Kenney04, Kenney14, Vollmer01}.

However, the recent discoveries of isolated\footnote{Typically defined as a dwarf galaxy located at least 500 kpc from a massive galaxy, with a line-of-sight relative velocity greater than 500 \kms \citep{Paudel14,Fong20,Bidaran25}.} dEs, which do not belong to any galaxy cluster or group, have reignited the debate about dE formation---whether all dEs necessarily form through the influence of their local environment \citep{Janz17, Bidaran25}. Although dwarf--dwarf galaxy mergers do occur in low-density environments \citep{Paudel18}, they tend to involve at least one star-forming dwarf galaxy.

From a theoretical perspective, the existence of backsplash galaxies---systems that once resided within the virial radius of a galaxy cluster or group but have since traveled beyond it---has important implications for our understanding of galaxy evolution in different environments \citep{Gill05, Pimbblet11, Borrow23}. These galaxies provide evidence that environmental effects such as RPS, tidal interactions, and galaxy harassment are not strictly confined within cluster boundaries. Instead, galaxies can undergo significant environmental processing deep within a dense halo and later appear in regions traditionally considered the domain of isolated systems \citep{Wetzel14}. This challenges the classical dichotomy between ``field'' and ``cluster'' galaxies and suggests that a galaxy’s present-day location does not necessarily reflect its full interaction history.

\begin{figure}
\includegraphics[width=8.5cm]{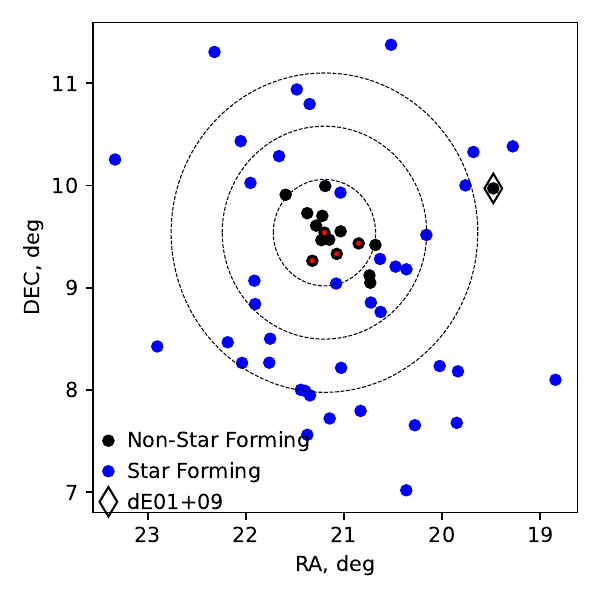}
\includegraphics[width=8.5cm]{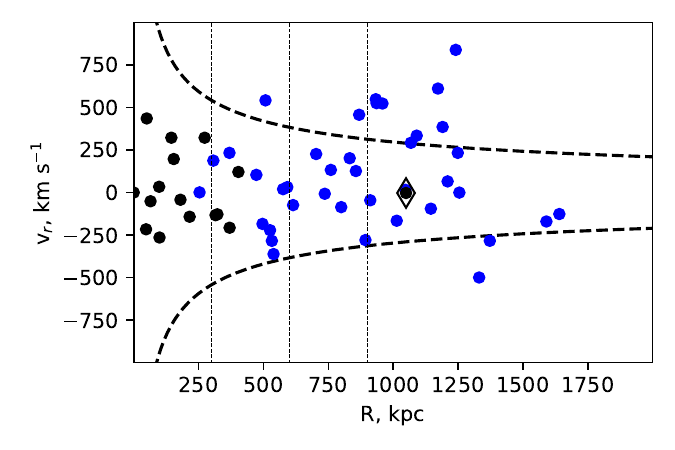}
\caption{\footnotesize 
Top: Spatial distribution of galaxies around NGC\,524 from NED, within a velocity range of $\pm$1000\,\kms relative to the line-of-sight velocity of NGC\,524. Star-forming and non-star-forming galaxies are shown as black and blue, respectively. The red dot represents massive (M$_{*} > 10^{10}$\,\,M$_{\sun}$) galaxies. The three concentric black short-dashed circles indicate projected distances of one, two, and three times the group’s virial radius. The position of \isde is highlighted with an open diamond symbol. 
Bottom: The phase-space diagram of galaxies around NGC\,524. The thick dashed loci indicate the escape velocity profile of the group, assuming a total mass of 10$^{13}$\,M$_{\sun}$. Vertical short-dashed lines denote one, two, and three times the group’s virial radius.
}
\label{sky}
\end{figure}

In particular, \citet{Chilingarian15} argued that some observed isolated compact elliptical (cE) galaxies and quiescent dwarf galaxies---systems found far from any massive host---may not have formed {\it in situ}. Rather, they could be the remnants of galaxies that were tidally stripped during close passages through a group or cluster and subsequently ejected to the outskirts or beyond the virial radius. These ``runaway'' galaxies therefore occupy regions that appear isolated today but carry the signatures of a turbulent past in denser environments.

In this study, we report the discovery of a dE, SDSS J011754.86+095819.0 (hereafter \isde), which appears to be located at a large distance from its host group. It currently resides in a nearly isolated region, at a projected distance of approximately one megaparsec from its likely host group, NGC\,524.\footnote{This distance is based on the assumed distance to NGC\,524 of 34.2\,Mpc from us, derived from the Hubble flow using a cosmology with $H_{0}= 71$ and  $\Omega_{m} = 0.3$.}

\section{Finding isolated early-type dwarf galaxies}
We have conducted a systematic search for low-mass early-type galaxies across various environments---clusters, groups, and fields---within the local volume (z\,$<$\,0.01). To achieve this, we visually inspected a large area of the sky using data from the SDSS and Legacy imaging surveys, identifying 5,054 dEs located in clusters and groups. The resulting main catalog was published in \citet{Paudel23}. The strength of this catalog lies in its detailed description of the spatial distribution of dEs in relation to their local environments. Our findings indicate that a significant majority of present-day dEs are found in clusters and become increasingly rare in low-density regions. Additionally, we observed a correlation between dEs and their nearest bright neighbor galaxies, suggesting that dEs are more likely to form in regions where their nearest bright neighbors are non-star-forming.

To advance our efforts in identifying dEs in low-density environments, we analyze the morphological properties of dwarf galaxies observed in the DESI survey (DR1) \citep{DESI25}. By combining the redshift catalogs from SDSS and DESI, we characterize the local environments of dwarf galaxies within $z\,<\,0.01$. For morphological classification, we employ a state-of-the-art machine learning tool trained on a reference sample of 5,000 previously cataloged dEs (see Appendix for details). This analysis leads to the identification of 751 dEs with confirmed redshifts, including 183 not found in earlier catalogs. The full methodology and results will be presented in an upcoming publication. Notably, one of these (\isde) lies in near-complete isolation, making it particularly intriguing.

\isde is located at a redshift of $z\,=\,0.008$ to the east of the NGC\,524 group, at an angular distance of 1.2 degrees---corresponding to approximately 1.2 Mpc in sky-projected physical scale---from the group center. The top panel of Figure \ref{sky} shows the distribution of galaxies around the group within a line-of-sight radial velocity range of $\pm$1000\,\kms relative to NGC\,524. Within NGC\,524, most member galaxies are indeed non-star-forming within the virial radius of 300 kpc.\footnote{The virial radius is calculated for a halo mass of 10$^{13}$\,M$_{\sun}$, which we derived using the projected group mass equation provided by \citet{Heisler85}.} Unexpectedly, we find a non-star-forming dwarf, \isde, located beyond three times the virial radius from the center of the group. Interestingly, we do not find any massive galaxies with $M_{*}\,>\,10^{10}\,M_{\sun}$ within a sky-projected radius of 700\,kpc and a radial velocity range of $\pm$500\,\kms from \isde. In this sense, \isde can be considered a truly isolated dE, as many previous studies have used this criterion to identify isolated dwarf galaxies \citep{Fong20, Bidaran25}. Interestingly, as shown in the bottom panel of the figure, \isde shares a similar radial velocity with the central galaxy of the group, yet it is situated in a region devoid of other non-star-forming galaxies. The presence of a quiescent dwarf in such an isolated environment is both unusual and noteworthy.

\begin{table}
\footnotesize
\caption{\small The properties of \isde. }
\begin{tabular}{lc r}
\hline
Parameter & Value & Remark \\
\hline
RA (hh:mm:ss) & 01:17:17:54.85 & 19.478555 deg \\
Dec (dd:mm:ss) & 09:58:19.10 & +9.971973 deg \\
$D_{ proj}$ &  1.8 deg (1.1 Mpc) & from\,NGC\,524\\
v$_{r}$ & 2400$\pm$15\,\kms & reported by DESI \\
v$_{r}$ & 2371$\pm$29\,\kms & measured by us \\
$M_{r}$ & $-$15.72$\pm$0.01 mag & \\
$g-r$ & 0.67$\pm$0.02 mag & \\
$M_{*}$ & 2.8$\times$10$^{8}$\,M$_{\sun}$ &  \\
\hline
$R_{e}$  & 1.2$\pm$0.2 kpc & Global\\
$n$    & 1.1$\pm$0.1  & Global \\
$\mu_{e,g}$ & 23.4$\pm$0.2  mag arcsec$^{-2}$ & Global\\
\hline
Age  & 8.3$\pm$1.4 Gyr & SSP\\
Z      & $-$1.19$\pm$0.21 dex & SSP\\
\hline
\end{tabular}
\tablefoot{The stellar mass is estimated from the $r$-band luminosity using a mass-to-light ratio calibrated from the $g-r$ color, following the empirical relation of \citet{Zhang17}. Structural parameters are derived from surface photometry performed on CFHT $g$-band images. Stellar population properties are obtained via full-spectrum fitting of the DESI optical spectrum, employing E-MILES SSP models.
}
\end{table}

\begin{figure}
\includegraphics[width=9.1cm]{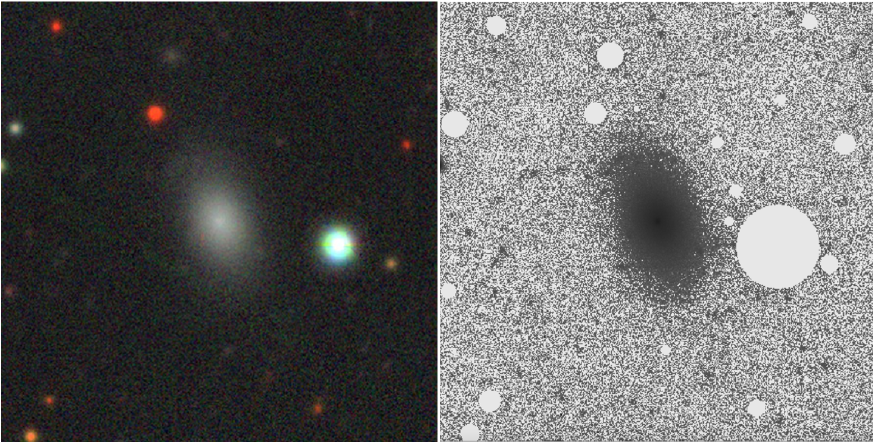}
\caption{\footnotesize  \isde as seen in Legacy Survey images. The left panel shows a $g-r-z$ combined color image obtained from the Legacy Survey viewer tool. The right panel displays a grayscale CFHT $g$-band image with a field of view of 1.5\arcmin\,$\times$\,1.5\arcmin. The color stretching is chosen to best reveal low-surface-brightness features, and unrelated background and foreground objects have been masked.
}
\label{main}
\end{figure}

\section{Data Analysis}
\subsection{Imaging}
A view of \isde in optical images is shown in Figure~\ref{main}. The left panel presents a $g$–$r$–$z$ color composite from the Legacy Survey. To better examine the low-surface-brightness component, we retrieved $g$-band CFHT/MegaCam data (500,s total exposure) from the CADC archive, which is shown in the left panel. The pre-processed Level 2 images, with 0.9$\arcsec$ seeing and 0.18$\arcsec$\,pix$^{-1}$ sampling, offer significantly higher quality than the Legacy data. As the target lies entirely on a single CCD chip, sky background subtraction—performed following \citet{Paudel18}—is straightforward despite the known mosaic-related issues with CFHT images.

\begin{figure}
\includegraphics[width=8.5cm]{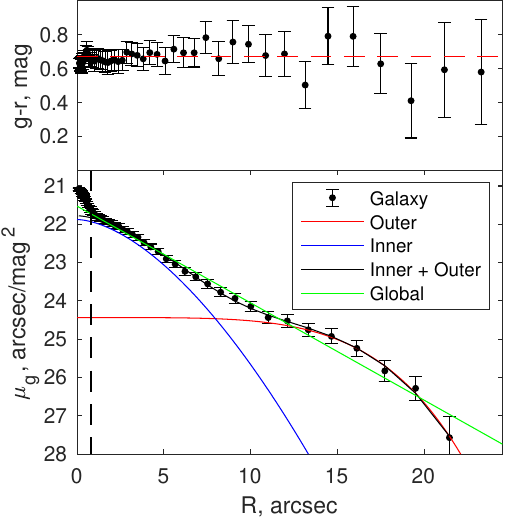}
\caption{\footnotesize Top: The $g-r$ color profile of \isde along its major axis. The red dashed horizontal line indicates the overall $g-r$ color. Bottom: $g$‑band surface‑brightness profile of \isde\ measured along its major axis. The solid blue line represents the best‑fitted inner S\'ersic component, while the red line shows the outer S\'ersic component, and their combined is plotted in black. The overall global best-fitted profile is shown in green. Uncertainties on the data points are indicated by the error bars. The vertical dashed line represents the radius of the DESI fibre.
}
\label{prof}
\end{figure}

We perform surface photometry on the CFHT MegaCam $g$-band image of \isde using the IRAF \texttt{ellipse} task \citep{Jedrzejewski87}. Before fitting, we subtract the sky background following the method in \citet{Paudel23} and manually mask unrelated foreground and background objects. Elliptical isophotes are fitted with a fixed center and position angle, while allowing the ellipticity to vary. The center of the galaxy is determined using \texttt{imcntr}, and the input parameters were refined through iterative runs of \texttt{ellipse}.

In the top panel of Figure~\ref{prof}, we present the $g-r$ color profile of \isde. The profile remains nearly flat within the uncertainties, with an average color of $g-r \simeq 0.67$ mag, as indicated by the horizontal dashed line. This uniform color suggests a relatively homogeneous stellar population throughout the galaxy and shows no evidence of recent central star formation. Such a flat profile contrasts with those of dEs located in the outskirts of the Virgo Cluster, where centrally concentrated star formation is commonly observed \citep{Lisker08,Urich17}.  The lower panel of Figure~\ref{prof} presents the $g$-band major-axis light profile of \isde, which shows a distinct break at $\sim$12$\arcsec$, suggesting a possible multi-component structure. A two-component S\'ersic fit yields $n_{in}=0.7$ and $n_{out}=0.4$ with effective radii $R_{e,{in}}=5\arcsec$ and $R_{e,{out}}=10\arcsec$, respectively. For the global structural parameters, we fit a single S\'ersic model to the entire light profile (excluding the central 2$\arcsec$ to avoid contamination from the nuclear light peak), obtaining $n=1.1$ and $R_{e}=1.2$\,kpc. These values are typical for dE galaxies with $M_{g}\simeq-15.5$\,mag \citep{Janz14}.

\subsection{Spectroscopy}
We retrieve the optical spectrum of \isde from the DESI archive. As expected, the DESI fiber spectrum shows no Balmer emission lines. To maximize the information extracted, we perform full-spectrum fitting using the $pPXF$ package \citep{Cappellari04, Cappellari23}, which employs a penalized likelihood method to match the observed spectrum with combinations of simple stellar populations (SSPs) defined by age and metallicity. We use templates from the E-MILES library \citep{Vazdekis16}, restricting the fit to wavelengths below 5700\AA{} to minimize the sensitivity to the initial mass function. Given the relatively low signal-to-noise ratio, we smooth the spectrum with a three-pixel Gaussian kernel before fitting\footnote{The DESI blue-channel spectrum has an average resolving power of  $R\,\approx\,2000$ (FWHM $\approx$ 1.8\,\text{\AA}) with a sampling of 0.8\,\text{\AA}\,pix$^{-1}$. Applying a three-pixel Gaussian  smoothing yields a final resolution of $R\,\approx\,1746$  (FWHM $\approx$2.9\,\text{\AA}) at a rest-frame wavelength of  5000\,\text{\AA}.}

\begin{figure}
\includegraphics[width=8.7cm]{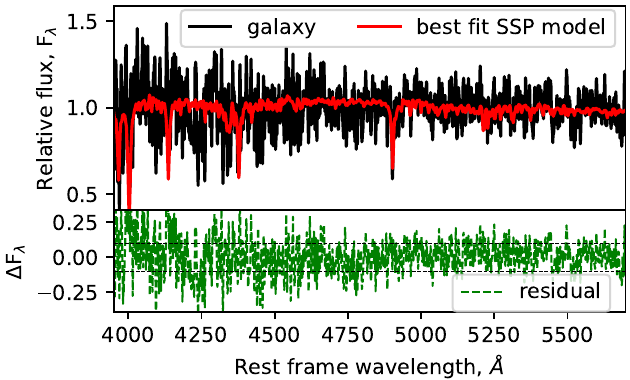}
\caption{\footnotesize Top: The DESI fiber spectrum of \isde (black) overlaid with the best-fitting simple stellar population (SSP) model spectrum (red). Bottom: The residuals (green), defined as the difference between the observed and model spectra. The fit reproduces the observed flux to within approximately 10\,\%, as indicated by the horizontal dashed lines.
}
\label{spec}
\end{figure}

The quality of the model fit to the DESI spectrum is shown in Figure \ref{spec}. The SSP age and metallicity (log[Z/Z$_{\sun}$]) of \isde, derived from the DESI fiber spectroscopy, are 8.3$\pm$1.4\,Gyr and $-$1.18$\pm$0.21\,dex, respectively.
The best fitted radial velocity we obtained is 2374$\pm$29\,\kms, which is similar to the reported value of radial velocity of \isde by the DESI DR1 catalog. It should be noted that the DESI fiber spectrum samples only the central 1.5$\arcsec$ region of the galaxy; therefore, the derived stellar population properties pertain to its central region. Following the approach recommended by \citet{Cappellari23}, we estimated the uncertainties in age and metallicity using a bootstrap analysis, similar to the method of \citet{Kacharov18} and as described on the $pPXF$ example pages. For each science spectrum, we generated 100 Monte Carlo realizations by adding noise drawn from the residuals of the best-fitting model. Each realization was refitted with the same pPXF setup, and the resulting distributions of light-weighted age and metallicity were used to determine the corresponding $1\sigma$ uncertainties.

\section{Conclusions and Discussion}
We have identified a rare dE, \isde, located in a nearly isolated environment at a sky-projected distance of 1.2\,Mpc from its nearest group, NGC\,524. \isde exhibits typical morphological and stellar population characteristics of a dE. 
Given the current data and the rarity of low-mass quenched galaxies like \isde in the field, we argue that \isde is more likely a runaway or backsplash candidate than a true field galaxy.

In this scenario, dE01+09 may have entered the group several gigayears ago as a star-forming dwarf, where it experienced environmental quenching --likely via a combination of ram-pressure stripping and tidal interactions—approximately 8.3 Gyr ago, as we expect from its SSP age. After quenching, it continued to orbit within the group potential for several gigayears. Roughly 3.5 Gyr ago, a strong dynamical interaction, possibly involving a three-body encounter, imparted a velocity near the group’s escape speed ($v_{\rm esc}$\,$\simeq$\,300\,\kms in the plane of the sky) and ejected the galaxy beyond the virial radius. At the present epoch, dE01+09 lies at least 1.2 Mpc in projection from the group centre, is quiescent, and may be deficient in dark matter, as suggested by analogous cases in simulations \citet{Ana23}.

While high-velocity ejections are more common in massive clusters (M $\approx$ $10^{14} - 10^{15}$\,M$_{\sun}$), numerical studies demonstrate that backsplash and even unbound orbits also occur in group environments (M $\approx10^{13}$ M$_{\sun}$) when orbital resonances or close multi-body encounters transfer sufficient energy to satellites \citep{Wetzel14,Levis25}. The escape velocity of the NGC 524 group ($\sim$300\,\kms) is comparable to that required to unbind a low-mass dwarf, and thus makes the proposed scenario dynamically plausible despite the group’s lower velocity dispersion. Indeed, this scenario is limited by the lack of direct constraints on the three-dimensional positions and velocities of dE01+09 and NGC\,524, making the ejection timescale and trajectory uncertain.

Our findings align with results from \cite{Benavides21}, who used the TNG50 simulation to trace backsplash galaxies in both groups and clusters. They find that in groups, quenched ultra diffuse galaxies with stellar masses similar to dE01+09 can be found as far as $\sim$3\,R$_{2000}$ from the group or cluster centre, with a smaller fraction reaching $\sim$3.5\,R$_{2000}$—the current projected separation of our target. This suggests that dE01+09 represents an extreme but still plausible outcome of group dynamical evolution.

\newpage

\begin{appendix}

\section{Image classification}

\begin{figure}[h]
\includegraphics[width=8.5cm]{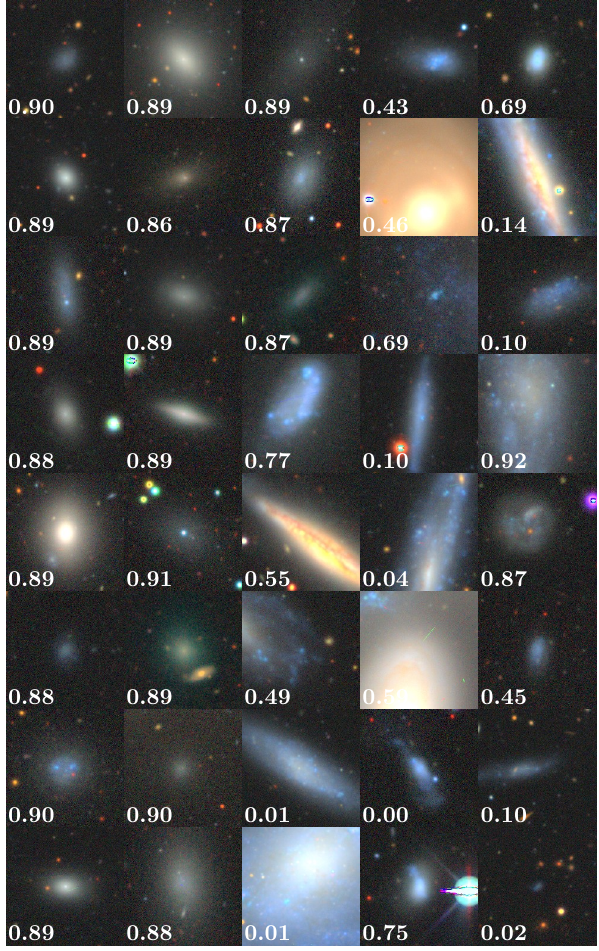}
\caption{\footnotesize Classification of galaxies using the retrained MobileNetV2 model. Each panel shows a galaxy image along with its predicted probability of being a dE. As expected, the model assigns the highest probabilities to galaxies with the smooth morphological characteristics typical of dEs, while the probability decreases for galaxies whose visual appearance diverges from these features.
}
\label{declass}
\end{figure}

To identify dEs in large imaging surveys, we employ MobileNetV2, a lightweight convolutional neural network (CNN) optimized for efficient image classification. We retrain the model using our previously compiled catalog of 5,000 visually confirmed dEs as the training set. For the input data, we generate image cutouts from the Legacy Survey in the $gri$ bands, centered on each galaxy and resized to 300\,$\times$\,300 pixels to capture relevant morphological features while maintaining computational efficiency. The retrained MobileNetV2 achieves a classification accuracy of 87\,\%, with a precision of 83\,\% and a recall of 79\,\% for the dE class on a held-out validation set.

The trained MobileNetV2 model is then applied to classify a new sample of 22,000 galaxies at redshift $z$\,$<$\,0.01, drawn from the DESI DR1 imaging catalog. During inference, the model assigns probabilities for each galaxy belonging to the dE class based on features such as a smooth light distribution, compactness, and the absence of star-forming structures. An example output from our automated classification is shown in Figure \ref{declass}. The complete analysis, along with a full catalog of newly spectroscopically confirmed dEs, will be presented in a forthcoming publication.

\section{Acknowledgements}
\begin{acknowledgements}
SP and SJY acknowledge support from the Mid-career Researcher Program (RS-2023-00208957 and RS-2024-00344283, respectively) through Korea's National Research Foundation (NRF). 
SJY and CGS acknowledge support from the Basic Science Research Program (2022R1A6A1A03053472 and 2018R1A6A1A06024977, respectively) through Korea's NRF funded by the Ministry of Education. 

The DESI Legacy Imaging Surveys consist of three individual and complementary projects: the Dark Energy Camera Legacy Survey (DECaLS), the Beijing-Arizona Sky Survey (BASS), and the Mayall z-band Legacy Survey (MzLS). DECaLS, BASS and MzLS together include data obtained, respectively, at the Blanco telescope, Cerro Tololo Inter-American Observatory, NSF’s NOIRLab; the Bok telescope, Steward Observatory, University of Arizona; and the Mayall telescope, Kitt Peak National Observatory, NOIRLab. NOIRLab is operated by the Association of Universities for Research in Astronomy (AURA) under a cooperative agreement with the National Science Foundation. Pipeline processing and analyses of the data were supported by NOIRLab and the Lawrence Berkeley National Laboratory (LBNL). Legacy Surveys also uses data products from the Near-Earth Object Wide-field Infrared Survey Explorer (NEOWISE), a project of the Jet Propulsion Laboratory/California Institute of Technology, funded by the National Aeronautics and Space Administration. Legacy Surveys was supported by: the Director, Office of Science, Office of High Energy Physics of the U.S. Department of Energy; the National Energy Research Scientific Computing Center, a DOE Office of Science User Facility; the U.S. National Science Foundation, Division of Astronomical Sciences; the National Astronomical Observatories of China, the Chinese Academy of Sciences and the Chinese National Natural Science Foundation. LBNL is managed by the Regents of the University of California under contract to the U.S. Department of Energy. The complete acknowledgments can be found at https://www.legacysurvey.org/acknowledgment/.

This research used data obtained with the Dark Energy Spectroscopic Instrument (DESI). DESI construction and operations is managed by the Lawrence Berkeley National Laboratory. This material is based upon work supported by the U.S. Department of Energy, Office of Science, Office of High-Energy Physics, under Contract No. DE–AC02–05CH11231, and by the National Energy Research Scientific Computing Center, a DOE Office of Science User Facility under the same contract. Additional support for DESI was provided by the U.S. National Science Foundation (NSF), Division of Astronomical Sciences under Contract No. AST-0950945 to the NSF’s National Optical-Infrared Astronomy Research Laboratory; the Science and Technology Facilities Council of the United Kingdom; the Gordon and Betty Moore Foundation; the Heising-Simons Foundation; the French Alternative Energies and Atomic Energy Commission (CEA); the National Council of Science and Technology of Mexico (CONACYT); the Ministry of Science and Innovation of Spain (MICINN), and by the DESI Member Institutions: www.desi.lbl.gov/collaborating-institutions. The DESI collaboration is honored to be permitted to conduct scientific research on Iolkam Du’ag (Kitt Peak), a mountain with particular significance to the Tohono O’odham Nation. Any opinions, findings, and conclusions or recommendations expressed in this material are those of the author(s) and do not necessarily reflect the views of the U.S. National Science Foundation, the U.S. Department of Energy, or any of the listed funding agencies.

\end{acknowledgements}

\end{appendix}


\begin{thebibliography}{40}
\expandafter\ifx\csname natexlab\endcsname\relax\def\natexlab#1{#1}\fi

\bibitem[{{Benavides} {et~al.}(2021){Benavides}, {Sales}, {Abadi}, {Pillepich},
  {Nelson}, {Marinacci}, {Cooper}, {Pakmor}, {Torrey}, {Vogelsberger}, \&
  {Hernquist}}]{Benavides21}
{Benavides}, J.~A., {Sales}, L.~V., {Abadi}, M.~G., {et~al.} 2021, Nature
  Astronomy, 5, 1255

\bibitem[{{Bidaran} {et~al.}(2025){Bidaran}, {P{\'e}rez},
  {S{\'a}nchez-Menguiano}, {Argudo-Fern{\'a}ndez}, {Ferr{\'e}-Mateu},
  {Navarro}, {Peletier}, {Ruiz-Lara}, {van de Ven}, {Verley}, {Zurita}, {Duarte
  Puertas}, {Falc{\'o}n-Barroso}, {S{\'a}nchez-Bl{\'a}zquez}, \&
  {Jim{\'e}nez}}]{Bidaran25}
{Bidaran}, B., {P{\'e}rez}, I., {S{\'a}nchez-Menguiano}, L., {et~al.} 2025,
  \aap, 693, L16

\bibitem[{{Binggeli} {et~al.}(1988){Binggeli}, {Sandage}, \&
  {Tammann}}]{Binggeli88}
{Binggeli}, B., {Sandage}, A., \& {Tammann}, G.~A. 1988, \araa, 26, 509

\bibitem[{{Borrow} {et~al.}(2023){Borrow}, {Vogelsberger}, {O'Neil},
  {McDonald}, \& {Smith}}]{Borrow23}
{Borrow}, J., {Vogelsberger}, M., {O'Neil}, S., {McDonald}, M.~A., \& {Smith},
  A. 2023, \mnras, 520, 649

\bibitem[{{Boselli} \& {Gavazzi}(2006)}]{Boselli06}
{Boselli}, A. \& {Gavazzi}, G. 2006, \pasp, 118, 517

\bibitem[{{Cappellari}(2023)}]{Cappellari23}
{Cappellari}, M. 2023, \mnras, 526, 3273

\bibitem[{{Cappellari} \& {Emsellem}(2004)}]{Cappellari04}
{Cappellari}, M. \& {Emsellem}, E. 2004, \pasp, 116, 138

\bibitem[{{Chilingarian} \& {Zolotukhin}(2015)}]{Chilingarian15}
{Chilingarian}, I. \& {Zolotukhin}, I. 2015, Science, 348, 418

\bibitem[{{Cortese} {et~al.}(2021){Cortese}, {Catinella}, \&
  {Smith}}]{Cortese21}
{Cortese}, L., {Catinella}, B., \& {Smith}, R. 2021, \pasa, 38, e035

\bibitem[{{Croton} {et~al.}(2006){Croton}, {Springel}, {White}, {De Lucia},
  {Frenk}, {Gao}, {Jenkins}, {Kauffmann}, {Navarro}, \& {Yoshida}}]{Croton06}
{Croton}, D.~J., {Springel}, V., {White}, S. D.~M., {et~al.} 2006, \mnras, 365,
  11

\bibitem[{{DESI Collaboration} {et~al.}(2025){DESI Collaboration},
  {Abdul-Karim}, {Adame}, {Aguado}, {Aguilar}, {Ahlen}, {Alam}, {Aldering}, \&
  {Alexander}}]{DESI25}
{DESI Collaboration}, {Abdul-Karim}, M., {Adame}, A.~G., {et~al.} 2025, arXiv
  e-prints, arXiv:2503.14745

\bibitem[{{Gill} {et~al.}(2005){Gill}, {Knebe}, \& {Gibson}}]{Gill05}
{Gill}, S. P.~D., {Knebe}, A., \& {Gibson}, B.~K. 2005, \mnras, 356, 1327

\bibitem[{{Gunn} \& {Gott}(1972)}]{Gunn72}
{Gunn}, J.~E. \& {Gott}, J.~R.~I. 1972, \apj, 176, 1

\bibitem[{{Heisler} {et~al.}(1985){Heisler}, {Tremaine}, \&
  {Bahcall}}]{Heisler85}
{Heisler}, J., {Tremaine}, S., \& {Bahcall}, J.~N. 1985, \apj, 298, 8

\bibitem[{{Hopkins} {et~al.}(2012){Hopkins}, {Quataert}, \&
  {Murray}}]{Hopkins12}
{Hopkins}, P.~F., {Quataert}, E., \& {Murray}, N. 2012, \mnras, 421, 3522

\bibitem[{{Janz} {et~al.}(2014){Janz}, {Laurikainen}, {Lisker}, {Salo},
  {Peletier}, {Niemi}, {Toloba}, {Hensler}, {Falc{\'o}n-Barroso}, {Boselli},
  {den Brok}, {Hansson}, {Meyer}, {Ry{\'s}}, \& {Paudel}}]{Janz14}
{Janz}, J., {Laurikainen}, E., {Lisker}, T., {et~al.} 2014, \apj, 786, 105

\bibitem[{{Janz} {et~al.}(2017){Janz}, {Penny}, {Graham}, {Forbes}, \&
  {Davies}}]{Janz17}
{Janz}, J., {Penny}, S.~J., {Graham}, A.~W., {Forbes}, D.~A., \& {Davies},
  R.~L. 2017, \mnras, 468, 2850

\bibitem[{{Jedrzejewski}(1987)}]{Jedrzejewski87}
{Jedrzejewski}, R.~I. 1987, \mnras, 226, 747

\bibitem[{{Kacharov} {et~al.}(2018){Kacharov}, {Neumayer}, {Seth},
  {Cappellari}, {McDermid}, {Walcher}, \& {B{\"o}ker}}]{Kacharov18}
{Kacharov}, N., {Neumayer}, N., {Seth}, A.~C., {et~al.} 2018, \mnras, 480, 1973

\bibitem[{{Kado-Fong} {et~al.}(2020){Kado-Fong}, {Greene}, {Greco}, {Beaton},
  {Goulding}, {Johnson}, \& {Komiyama}}]{Fong20}
{Kado-Fong}, E., {Greene}, J.~E., {Greco}, J.~P., {et~al.} 2020, \aj, 159, 103

\bibitem[{{Kenney} {et~al.}(2014){Kenney}, {Geha}, {J{\'a}chym}, {Crowl},
  {Dague}, {Chung}, {van Gorkom}, \& {Vollmer}}]{Kenney14}
{Kenney}, J.~D.~P., {Geha}, M., {J{\'a}chym}, P., {et~al.} 2014, \apj, 780, 119

\bibitem[{{Kenney} {et~al.}(2004){Kenney}, {van Gorkom}, \&
  {Vollmer}}]{Kenney04}
{Kenney}, J.~D.~P., {van Gorkom}, J.~H., \& {Vollmer}, B. 2004, \aj, 127, 3361

\bibitem[{{Larson} {et~al.}(1980){Larson}, {Tinsley}, \& {Caldwell}}]{Larson80}
{Larson}, R.~B., {Tinsley}, B.~M., \& {Caldwell}, C.~N. 1980, \apj, 237, 692

\bibitem[{{Levis} {et~al.}(2025){Levis}, {Coenda}, {Muriel}, {de los Rios},
  {Ragone-Figueroa}, {Mart{\'\i}nez}, \& {Ruiz}}]{Levis25}
{Levis}, S., {Coenda}, V., {Muriel}, H., {et~al.} 2025, \aap, 698, A57

\bibitem[{{Lisker} {et~al.}(2008){Lisker}, {Grebel}, \& {Binggeli}}]{Lisker08}
{Lisker}, T., {Grebel}, E.~K., \& {Binggeli}, B. 2008, \aj, 135, 380

\bibitem[{{Lisker} {et~al.}(2007){Lisker}, {Grebel}, {Binggeli}, \&
  {Glatt}}]{Lisker07}
{Lisker}, T., {Grebel}, E.~K., {Binggeli}, B., \& {Glatt}, K. 2007, \apj, 660,
  1186

\bibitem[{{Mayer} {et~al.}(2001){Mayer}, {Governato}, {Colpi}, {Moore},
  {Quinn}, {Wadsley}, {Stadel}, \& {Lake}}]{Mayer01}
{Mayer}, L., {Governato}, F., {Colpi}, M., {et~al.} 2001, \apj, 559, 754

\bibitem[{{Mitra{\v{s}}inovi{\'c}} {et~al.}(2023){Mitra{\v{s}}inovi{\'c}},
  {Smole}, \& {Micic}}]{Ana23}
{Mitra{\v{s}}inovi{\'c}}, A., {Smole}, M., \& {Micic}, M. 2023, \aap, 680, L1

\bibitem[{{Moore} {et~al.}(1996){Moore}, {Katz}, {Lake}, {Dressler}, \&
  {Oemler}}]{Moore96}
{Moore}, B., {Katz}, N., {Lake}, G., {Dressler}, A., \& {Oemler}, A. 1996,
  \nat, 379, 613

\bibitem[{{Paudel} {et~al.}(2014){Paudel}, {Lisker}, {Hansson}, \&
  {Huxor}}]{Paudel14}
{Paudel}, S., {Lisker}, T., {Hansson}, K.~S.~A., \& {Huxor}, A.~P. 2014,
  \mnras, 443, 446

\bibitem[{{Paudel} {et~al.}(2018){Paudel}, {Smith}, {Yoon},
  {Calder{\'o}n-Castillo}, \& {Duc}}]{Paudel18}
{Paudel}, S., {Smith}, R., {Yoon}, S.~J., {Calder{\'o}n-Castillo}, P., \&
  {Duc}, P.-A. 2018, \apjs, 237, 36

\bibitem[{{Paudel} {et~al.}(2023){Paudel}, {Yoon}, {Yoo}, {Smith}, {Chhatkuli},
  {Kumar Bachchan}, {Aryal}, {Adhikari}, {Adhikari}, {Sedain}, {Sheikh},
  {Dhital}, {Giri}, \& {Baral}}]{Paudel23}
{Paudel}, S., {Yoon}, S.-J., {Yoo}, J., {et~al.} 2023, \apjs, 265, 57

\bibitem[{{Pimbblet}(2011)}]{Pimbblet11}
{Pimbblet}, K.~A. 2011, \mnras, 411, 2637

\bibitem[{{Simpson} {et~al.}(2018){Simpson}, {Grand}, {G{\'o}mez}, {Marinacci},
  {Pakmor}, {Springel}, {Campbell}, \& {Frenk}}]{Simpson18}
{Simpson}, C.~M., {Grand}, R. J.~J., {G{\'o}mez}, F.~A., {et~al.} 2018, \mnras,
  478, 548

\bibitem[{{Smith} {et~al.}(2010){Smith}, {Davies}, \& {Nelson}}]{Smith10}
{Smith}, R., {Davies}, J.~I., \& {Nelson}, A.~H. 2010, \mnras, 405, 1723

\bibitem[{{Urich} {et~al.}(2017){Urich}, {Lisker}, {Janz}, {van de Ven},
  {Leaman}, {Boselli}, {Paudel}, {Sybilska}, {Peletier}, {den Brok}, {Hensler},
  {Toloba}, {Falc{\'o}n-Barroso}, \& {Niemi}}]{Urich17}
{Urich}, L., {Lisker}, T., {Janz}, J., {et~al.} 2017, \aap, 606, A135

\bibitem[{{Vazdekis} {et~al.}(2016){Vazdekis}, {Koleva}, {Ricciardelli},
  {R{\"o}ck}, \& {Falc{\'o}n-Barroso}}]{Vazdekis16}
{Vazdekis}, A., {Koleva}, M., {Ricciardelli}, E., {R{\"o}ck}, B., \&
  {Falc{\'o}n-Barroso}, J. 2016, \mnras, 463, 3409

\bibitem[{{Vollmer} {et~al.}(2001){Vollmer}, {Cayatte}, {Balkowski}, \&
  {Duschl}}]{Vollmer01}
{Vollmer}, B., {Cayatte}, V., {Balkowski}, C., \& {Duschl}, W.~J. 2001, \apj,
  561, 708

\bibitem[{{Wetzel} {et~al.}(2014){Wetzel}, {Tinker}, {Conroy}, \& {van den
  Bosch}}]{Wetzel14}
{Wetzel}, A.~R., {Tinker}, J.~L., {Conroy}, C., \& {van den Bosch}, F.~C. 2014,
  \mnras, 439, 2687

\bibitem[{{Zhang} \& {Bell}(2017)}]{Zhang17}
{Zhang}, Y. \& {Bell}, E.~F. 2017, \apjl, 835, L2

\end{thebibliography}
\end{document}